# Wide-angle Scanning Heterogeneous Element-Based Phased Array Using Novel Scanning Envelope Synthesis Method

Yongzheng Li, *Student Member, IEEE*, Wanchen Yang, *Senior Member, IEEE*, Quan Xue, *Fellow, IEEE* and Wenquan Che, *Fellow, IEEE*

*Abstract*—**Two novel methods, including the scanning envelope synthesis (SES) method and the active reflection self-cancellation (ARC) method, are proposed to design wide-angle scanning heterogeneous element phased arrays. Heterogeneous strategy is efficient to extend scanning range but quantitatively characterization of the effect is critically needed to guide design for achieving desired performance. The proposed SES method derives theoretically the relationship between scanning range and the 3dB-beamwidth of the pattern envelope of one phased array, which is linear superposition of active radiation pattern (AEP) magnitude of each element. Therefore, the contribution of each kind of heterogeneity can be quantitatively analyzed for further enhancing the scanning range. As we see, one high active reflection coefficient of the phased array can directly reduce the realized gain. In this way, one ARC method is proposed to reduce the active reflection coefficient by counteracting the reflection component of active reflection coefficient with its transmission component, thereby keeping the realized gain efficiently even when the array scans at large angels. For verification, one 24.5-29.5GHz 4x4 phased array scanning in E-plane is designed and fabricated. Benefiting from the proposed SES method, the scanning range of the prototype is extended up to ±74°, around 10° improvement over one traditional heterogeneous array. Meanwhile, the active reflection coefficient is reduced from -4dB to lower than -7.5dB by applying the ARC method.**

*Index Terms*—**Active reflection coefficient, beam scanning, heterogeneous elements, phased array, scanning envelope.**

## I. Introduction

Millimeter-wave communication, essential for large-capacity line-of-sight coverage applications, is a key component of B5G/6G. To meet the demands for coverage and capacity, millimeter-wave phased arrays must have wide-angle scanning capabilities.

Recently, many methods have been proposed to enhance the scanning range of a phased array. According to the array synthesis theory, scanning range can be improved by employing the wide beamwidth element. Several approaches have been demonstrated to achieve the wide beamwidth element, including introducing the in-phased mirror source [1-3], [5], exciting multimode [6], miniaturization of radiating aperture [7-8], exciting surface wave [9]. In [2], the 3-dB beamwidth was enhanced from 85° to 102° by constructing a vertical current perpendicular to the PEC ground. This approach achieved a scanning range of ±65° within a 12.7% bandwidth. The horizontal magnetic dipole was combined with AMC to realize a wide beamwidth element, and the scanning range covers ±90° in a bandwidth of 5%. However, additional structures need to be loaded to construct the in-phased mirror source, increasing the energy storage in an element, thus limiting the operating bandwidth. One zeroth-order resonance was excited in [6] combined with $TM_{010}$ mode to enhance the element's beamwidth. One scanning range of ±66° was achieved, but the bandwidth is limited by overlapping frequency bands of the two modes, which is only 4.09%. A shrined radiation aperture has the potential to enhance the beamwidth. For instance, the element was miniaturized in [7] and [8], achieving the scanning range up to ±60° and ±76°, respectively. However, the miniaturization of the radiation aperture also results in reduction of the impedance bandwidth. TM surface wave was excited in [9] to realize a wide scanning range up to ±81°. However, the impedance surface guiding the surface wave only operates within a narrow bandwidth of 3.6%, limiting the array bandwidth. Recently, One heterogeneous method based on the pseudo-conformal principle was proposed [10-12], capable of achieving wideband wide-angle scanning in phased arrays. Obviously, constructing heterogeneous elements with different beam directions transforms the aperture field into an imitation of a conformal surface array, thereby enhancing scanning range. However, the current research is limited to qualitative analysis, and the construction of heterogeneous elements is primarily confined to beam tilting to approximate the effect of conformal arrays. The lack of theoretical analysis places considerable constraints on the design of the elements. Accordingly, a quantitative methodology is required to facilitate the design of heterogeneous element arrays.

Furthermore, the active reflection coefficient represents an essential metric in phased array design, as a high reflection coefficient can result in a reduction in realized gain. In the wide-angle scanning phased arrays, the element spacing is close to half a wavelength, resulting in near 180° phase of

This work was supported in part by the National Natural Science Foundation of China under Grant 62321002 and Grant 61931009. *(Corresponding author: Wenquan Che)*

Y. Li, Q. Xue and W. Che are with the Guangdong Provincial Key Laboratory of Millimeter-Wave and Terahertz, Guangdong-Hong Kong-Macao Joint Laboratory for Millimeter-Wave and Terahertz, School of Electronic and Information Engineering, South China University of Technology, Guangzhou 510641, China. (e-mail: eeeelyz@mail.scut.edu.cn, eeqxue@scut.edu.cn, eewqche@scut.edu.cn).

W. Yang is with the College of Electronic and Information Engineering, Nanjing University of Aeronautics and Astronautics, Nanjing 211106, China, and also with the Guangdong Key Laboratory of Millimeter-Waves and Terahertz, the School of Electronic and Information Engineering, South China University of Technology, Guangzhou 510641, China. (e-mail: yangwanchen@126.com).

coupling coefficient between adjacent elements. However, when the array scans to large angles, excitation phase difference is also close to 180°. As a result, the coupling coefficients of each element tend to sum in phase, increasing the active reflection coefficient. Therefore, the realized gain is reduced and the scanning range is degraded. Loading decoupling structures [13] is one efficient approach to reduce the active reflection coefficient as it can reduce the transmission coefficient. However, the decoupling structures can introduce bandwidth limitations and may deteriorate antenna matching. Another approach to reduce the active reflection coefficient at large angles is loading wide-angle matching layers [14] to facilitate the transition from planar aperture to the space. However, the implementation of wide-angle matching layers in the millimeter-wave band is a challenge due to design complexity. Therefore, further research is needed to develop methods for reducing active reflection coefficient of phased arrays when scanning to large angles.

In this work, one SES method is proposed firstly to improve the scanning range of a heterogeneous element array. The SES method relates the scanning range with the pattern envelope, which is linear summation of elements' AEP magnitude. Thereby contribution of each kind of heterogeneity applied on element to scanning angle can be quantitatively evaluated. Therefore, the SES method provides one effective guidance to determine heterogeneity type applied on the element for further enhancing scanning range. In addition, one novel ARC method is proposed to alleviate the active reflection coefficient problem when the array scans to large angles. Specifically, by adjusting the reflection coefficient phase of the antenna elements, the active reflection coefficient can be significantly reduced due to the counteraction of its reflection and transmission components.

The manuscript is arranged as follows. Sec. II elaborates on the SES method and its application on the heterogeneous phased array's scanning angle improvement. In Sec. III, the ARC method is detailed and used to reduce the active reflection coefficient of the proposed heterogeneous phased array. Sec. IV presents the array measured results and some discussions. Finally, Sec. V gives the conclusion.

## II. SCANNING ENVELOPE SYNTHESIS METHOD AND ITS APPLICATION IN HETEROGENEOUS PHASED ARRAY DESIGN

To further improve the scanning angle of a heterogeneous array, a quantitative relationship between heterogeneities and scanning range is desirable. For such purpose, one scanning envelope synthesis (SES) method is proposed. The principle of the SES method is initially analyzed and then employed to improve the scanning range of a heterogeneous array.

### A. Theoretical Analysis of SES Method

As shown in Fig. 1, consider an N-element uniformly spaced linear array with equal amplitude and gradient phase excitation. Let the far-field active element pattern (AEP) of the $i^{th}$ element as $f_i(\theta)$. The gain of the array can be expressed as (1), where $k$ is the free-space wave number, $d$ is the array spacing, and $\Delta\varphi$ is the excitation phase difference between adjacent elements. The total incident power is denoted by $P_{inc}$, and $\eta_0$ is the free-space impedance. According to (1), for a given $\Delta\varphi$, the peak gain occurs at $\theta_{max} \approx \arcsin\left(\frac{\Delta\varphi}{kd}\right)$, with its value expressed in (2). During phased array scanning, $\Delta\varphi$ varies within $(-\pi, \pi)$, and each $\Delta\varphi$ corresponds to a point $(\theta_{max}, f_{max})$ on the angle-gain zone. Now, defining the gain envelope by sequentially connecting these points, and this curve can be expressed in (3).

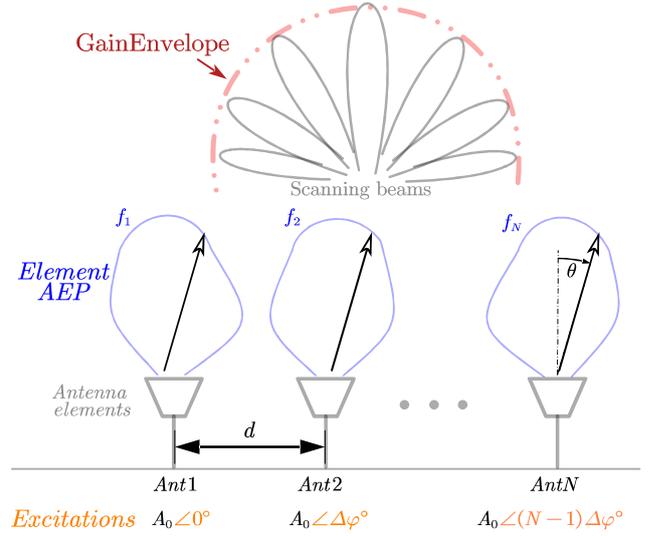

Fig. 1. Sketch of the N-element liner phased array.

$$G(\theta) = \frac{\left|\sum_n f_n(\theta) e^{-j(n-1)[kd\sin(\theta) - \Delta\varphi]}\right|^2}{P_{inc}\,\eta_0/2\pi} \quad \text{...................... (1)}$$

$$G_{max} = \frac{\left|\sum_n f_n(\arcsin(\Delta\varphi/kd))\right|^2}{P_{inc}\,\eta_0/2\pi} \quad \text{...................... (2)}$$

$$GainEnvelope(\theta) = \frac{\left|\sum_n f_n(\theta)\right|^2}{P_{inc}\,\eta_0/2\pi} \approx \frac{\left(\sum_n |f_n(\theta)|\right)^2}{P_{inc}\,\eta_0/2\pi} \quad \text{...... (3)}$$

$$PatternEnvelope(\theta) = \sum_n |f_n(\theta)| \quad \text{.................. (4)}$$

$$GainEnvelope(\theta) \approx \frac{PatternEnvelope(\theta)^2}{P_{inc}\,\eta_0/2\pi} \quad \text{............... (4.1)}$$

$$Scanning\ Range = width_{3dB}(dB(GainEnvelope(\theta))) \quad \text{....(4.2)}$$
$$\approx width_{3dB}(dB_{20}(PatternEnvelope(\theta)))$$

As we can see, when the far field reaches its peak at a specific spatial angle $\theta$, the radiation far field of each element $f_i(\theta)$ become nearly in phase. This allows the gain envelope to be approximated by square of the summation of each elements' AEP magnitude, as shown in (3). Now, defining the pattern envelope as in (4), thus the scanning gain envelope is related with square of the pattern envelope, as shown in (4.1). Meanwhile, as shown in (4.2), the scanning range is related with the 3dB-beamwidth of the scanning gain envelope, thus also related with the 3dB-beamwidth of the pattern envelope. Note that the pattern envelope is linear summation of magnitude of each elements' AEPs, which is easily calculated and can intuitively reflect the impact of AEP on scanning range. For example, as illustrated in Fig. 2, a more gradual decline in the pattern envelope is associated with a wider 3dB-beamwidth, thereby a wider scanning range. Therefore, carefully designing heterogeneous elements to generate suitable AEPs can achieve the desired wide-angle scanning range.

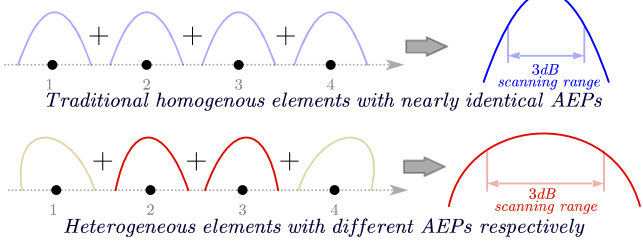

Fig. 2. Pattern envelope for phased arrays using traditional homogenous elements and heterogeneous elements.

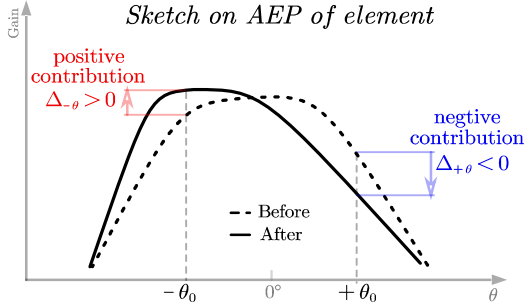

Fig. 3. Sketch on AEP of element before and after applying a specific heterogeneity.

$$PatternEnvelope(\theta) = Even(PatternEnvelope(\theta)) \dots \dots (5)$$

$$= 0.5\left(\sum_n |f_n(\theta)| + \sum_n |f_n(-\theta)|\right)$$

$$= \sum_n \frac{|f_n(\theta)| + |f_n(-\theta)|}{2} \dots \dots \dots (5.1)$$

$$Sf_n(\theta) = \frac{|f_n(\theta)| + |f_n(-\theta)|}{2} \dots \dots \dots (6)$$

$$PatternEnvelope(\theta) = \sum_n Sf_n(\theta) \dots \dots \dots (7)$$

$$\Delta_i = width_{3dB}(dB_{20}(Sf^{After})) - width_{3dB}(dB_{20}(Sf^{Before})) \dots (8)$$

Furthermore, due to the linear relationship between the pattern envelope and AEPs, the impact of each type of heterogeneity on the scanning range can be evaluated by analyzing the effects on the pattern envelope. For demonstration, after employing a specific heterogeneity on an array with symmetrical scanning performance, as shown in Fig. 3, the AEP magnitude increases by $\Delta_{-\theta_0}$ at $\theta = -\theta_0$ while decreasing by $\Delta_{+\theta_0}$ at $\theta = +\theta_0$. Therefore, the contribution to pattern envelope of using this heterogeneity at $\theta = \theta_0$ is calculated with $\Delta PatternEvenlope(\theta_0) = 0.5(\Delta_{-\theta_0} - \Delta_{+\theta_0})$. However, as we can see, the AEP variation at both $\pm\theta_0$ affect the pattern envelope at $\theta=\theta_0$, this introduces analysis ambiguity and inconvenience. To address this issue, even-odd decomposition is applied to (4). Meanwhile, the pattern envelope is essentially equal to its even component due to the symmetry in scanning performance. Therefore, substituting (4) into (5), the pattern envelope can be expressed with (5.1). Now, defining the symmetric AEP (S-AEP) in (6), (5.1) can thus be simplified as (7), which indicates that the pattern envelope is the summation of S-AEP of each element.

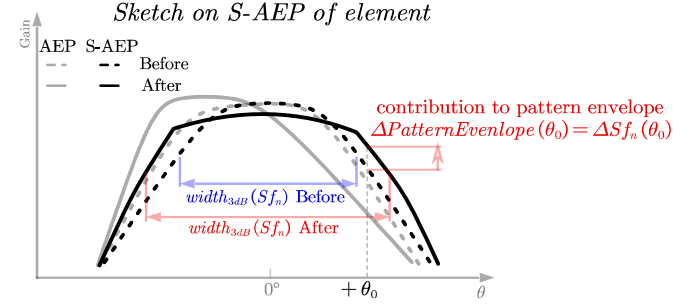

Fig. 4. Sketch on S-AEP of element before and after applying a specific heterogeneity.

The S-AEP features good characteristics that greatly facilitates the analyze of heterogeneity effect on scanning range. On the one hand, it reveals the effect of heterogeneity on pattern envelope more intuitively. As illustrated in Fig. 4, the S-AEP is an even function of $\theta$, and the variation produced by employing the heterogeneity in pattern envelope equals to the magnitude variation of S-AEP $\Delta Sf_n$. On the other hand, it also relates the heterogeneity effect with scanning range effectively. For a heterogeneity that positively contributes to scanning range, a wider 3dB-beamwidth in S-AEP can be observed, while the enhancement of S-AEP 3dB-beamwidth is proportional to the 3dB-beamwidth increment of the pattern envelope. As a wide scanning range corresponds to a flat pattern envelope, consequently, the difference in S-AEP 3dB-beamwidth before and after applying a heterogeneity, as defined in (8), serves as an intuitive indicator to quantify the contribution of a heterogeneity to the scanning range.

### B. Heterogeneous Phased Array Design and Verification of the SES Method

In the follows, the proposed SES method is validated and subsequently employed to investigate the heterogeneities effect on scanning range. For such purpose, two heterogeneous elements are initially designed. The evolution procedure of the heterogeneous element is illustrated in Fig. 5. *Element A* is a 2×2 metasurface antenna, and *Element B* is obtained from *Element A* by replacing the metasurface with a ladder-type one and loading a back cavity. Furthermore, *Element C* is realized by further loading two shorting pins asymmetrically. Note that all these elements are designed to operate within 24.5-29.5GHz. The renormalized radiation patterns of *Element A-C* at 29.5GHz are compared in Fig. 5(d). It can be observed that *Element B* exhibits an asymmetric pattern and a wider beamwidth compared to *Element A*, and this characteristic is further enhanced for *Element C*. Additionally, the beam direction for *Element B* and *C* deviates from broadside by 2° and 5°, respectively. Therefore, the employment of heterogeneity breaks symmetry of the radiation pattern and enhance its beamwidth, potentially altering the shape of the AEP in the array.

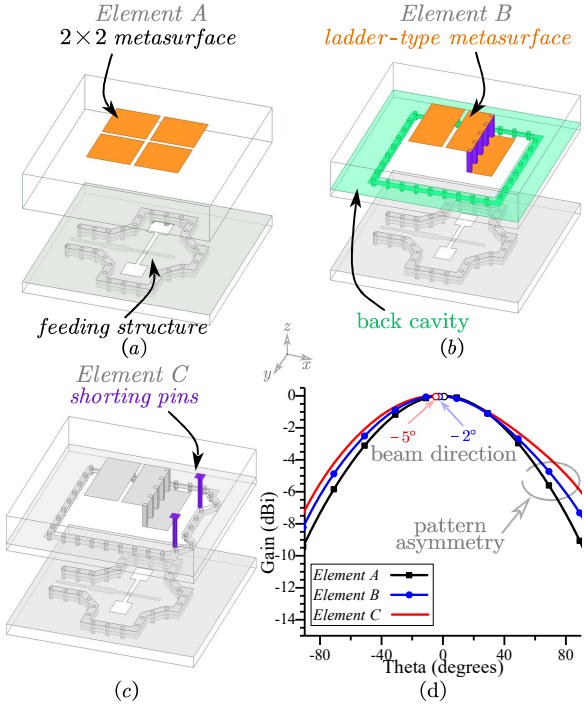

Fig. 5. Configuration for (a) *Element A*, (b) *Element B* and (c) *Element C*. (d) Renormalized radiation pattern in *xoz* plane for *Element A-C* at 29.5GHz.

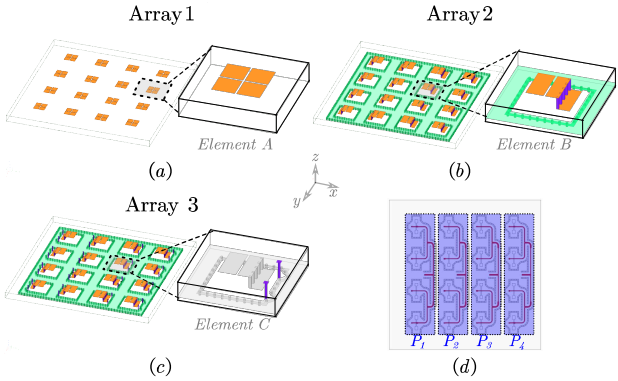

Fig. 6. Configuration of the phased arrays: (a) *Array 1*, (b) *Array* 2, (c) *Array 3*, (d) Configuration of the feeding network of these arrays.

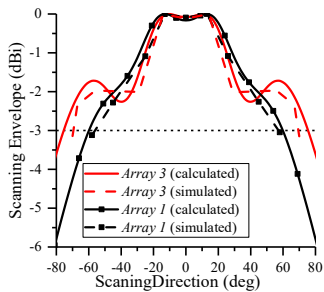

Fig. 7. Calculated and simulated scanning envelope (renormalized) for *Array* 1 and *Array* 3 at 29.5GHz.

Then several phased arrays scanning in the *xoz* plane are constructed using the *Element A-Element C*, termed with *Array*

1-*Array 3*, respectively. The configurations of these arrays are shown in Fig. 6(a-c), note that elements in *Array* 2- *Array* 3 are arranged symmetrically along the x direction to achieve a symmetric scanning range. An array spacing of 4.5mm is applied along both the x and y directions. All these arrays utilize the same feeding network, as illustrated in Fig. 6(d), which contains four ports $P_1$-$P_4$ arranged in the *xoz* plane. Each port is connected to a 1-to-4 power divider, which in turn feeds the elements in the *yoz* plane in phase.

To validate the accuracy of the gain envelope calculated in the SES method, the renormalized gain envelope at 29.5 GHz calculated from the AEP of elements is compared with the full-wave simulated results. As shown in Fig. 7, the gain difference between the calculated and simulated envelope is less than 0.5dB, indicating that the in-phase approximation holds and the SES method provides sufficient accuracy in predicting scanning envelope. Furthermore, it is observed that the *Array 3* exhibits a wider 3dB-beamwidth of gain envelope when compared with the homogeneous one (*Array* 1), indicating that the employment of heterogeneous elements enhances the scanning range.

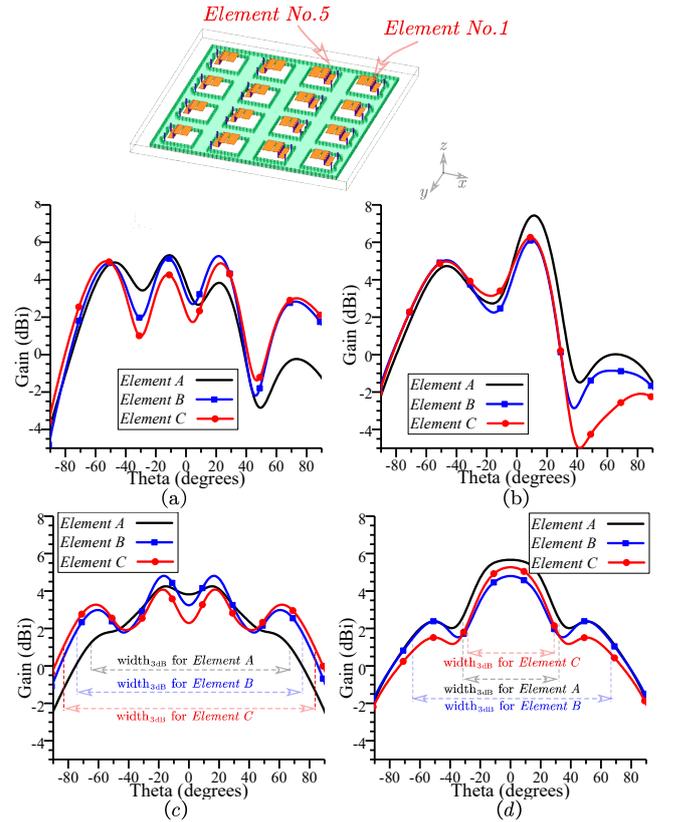

Fig. 8. AEPs of *Element A-C* at varies positon in array: (a) *Element No. 1*, (b) *Element No. 5*; S-AEPs of *Element A-C* at varies positon in array: (c) *Element No. 1*, (d) *Element No. 5*.

To gain a comprehensive understanding of how heterogeneities impact scanning range, a detailed examination from an element-specific perspective is conducted. For illustration, the AEP magnitude (in dB) at 29.5GHz in *xoz* plane for *Element No. 1* and *No. 5* in *Array*1-3 are compared in Fig. 8. It is observed that the employment of a ladder-typed metasurface and back cavity (*Element B* V.S. *Element A*) increases the magnitude of the AEP spanning -90° to -49° and 2° to 90° for *Element No. 1*, besides, the magnitude of the AEP is enhanced within the range of -90° to -25° for *Element No. 5*. Furthermore, the loading of asymmetric shorting pins (*Element C* V.S. *Element B*) enhances the magnitude of the AEP within the range of -90° to -51°, 28° to 90° for *Element No. 1*, and within the range of -35° to 15° for *Element No. 5*. The influence of heterogeneities on AEPs is complex, making it challenging to clearly determine their overall effect on scanning range.

Therefore, the S-AEPs are calculated and their 3dB-beamwidth are compared to quantitative analyses the influence of heterogeneities on scanning range. As illustrated in Fig. 8(c-d), *Element B* exhibit wider S-AEP 3dB-beamwidth than *Element A* for both *Element No.1* and *No. 5*, indicating that loading the ladder-type metasurface and back cavity improves the scanning range within 3dB gain variation. Meanwhile, the S-AEP 3dB-beamwidth of *Element C* is wider than that of *Element B* for the *Element No. 1*, while it is narrower for the *Element No. 5*. This indicates that loading the asymmetric shorting pins has different effect to scanning range depending on element position. For the elements located at the edge of the array in the E-plane, loading asymmetric shorting pins improves scanning range. However, for elements positioned at the inner side of the array in the E-plane, this loading has a detrimental effect on scanning range.

## C. Scanning Range Improvement of the Heterogeneous Phased Array within the Entire Frequency Band.

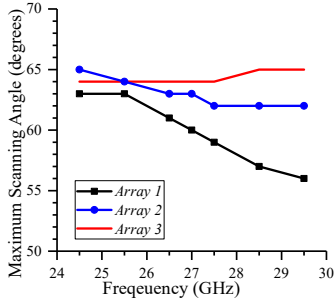

Fig. 9. Scanning range within 3-dB gain fluctuation vs. frequency for *Array* 1-3.

$$Sf_{Type\,i}(\theta) = \sum_{j \in \Omega_i} Sf_j(\theta) \quad\text{...................................................}\quad (9)$$

where: $\quad i \in \{I, II\}$

$$\Omega_I = \{1, 2, 3, 4\};$$
$$\Omega_{II} = \{5, 6, 7, 8\}.$$

$$\Delta_i = width_{3dB}\left(dB_{20}(Sf_{Type\,i}^{After})\right) - width_{3dB}\left(dB_{20}(Sf_{Type\,i}^{Before})\right) \quad\text{......}\quad (10)$$

To investigate the effect of these heterogeneities on scanning range within the whole band, the maximum scanning angle within 3-dB gain fluctuation across the entire frequency band is compared for *Array*1-*Array*3. As shown in Fig. 9, the employment of ladder-type metasurface and back-cavity can enhance scanning range by 2°-6°, with this enhancement gradually increase with frequency. Moreover, introducing asymmetric shorting pins results in scanning range improvement at frequency higher than 26GHz. As observed, the effect of heterogeneity on scanning envelope also varies with frequency. Therefore, to enhance the scanning range of the array within the frequency band, heterogeneities' contribution needs to be quantitatively evaluated at each frequency.

Since heterogeneities influence elements differently based on their position, for convenience of analyze, the array elements are divided into two groups for separate analysis. As depicted in Fig. 10(a), *Type I* are elements located near the edge of the array in the E-plane, while *Type II* are elements positioned at the inner side of the array. The contribution of each kind of heterogeneity on the pattern envelope for these two types of elements are required to be evaluated separately. To this end, the S-AEP for sub-arrays composed by each type of elements can be defined by (9). Likewise, the 3dB-beamwidth difference $\Delta_i$, originally defined in (8), can be extended as (10) to quantify the impact on scanning range.

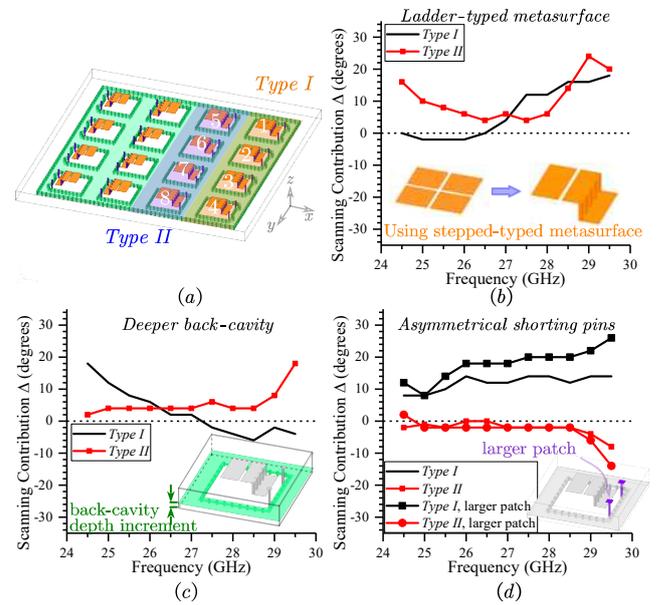

Fig. 10. (a) Classification of elements in array. Contribution of (b) using ladder-typed metasurface, (c) Increasing back-cavity depth and (d) loading asymmetrical shorting pins to scanning envelope.

The influence of adopting ladder-type metasurface on the scan envelope is investigated in Fig. 10(b). It is observed that for *Type II* elements, the ladder-type metasurface enhances scanning range across the entire frequency band, with greater contributions at two extremes frequency compared to the middle. Meanwhile, for *Type I* elements, there is a slight negative impact from 24.5 GHz to 26.5 GHz, but from 27 GHz to 29.5 GHz, the contribution turns positive and increases with frequency. The influence of back-cavity depth on scanning envelope is shown in Fig. 10(c). For *Type I* elements, increasing cavity depth has positively effect on scanning range below 27 GHz, while it negatively impacts higher frequencies. Conversely, increasing cavity depth of *Type II* elements consistently show improved scanning range across the entire frequency band, with increasing contribution at higher frequencies. Fig. 10(d) illustrates the effect of loading asymmetric shorting pins on the scanning envelope. This heterogeneity consistently enhances the scanning range for *Type I* elements, with greater benefits at higher frequencies. However, for *Type II* elements, the contribution of loading asymmetric shorting pins is minimal or negative. Additionally, the aforementioned effects are amplified by increasing the width of patch loaded on top of the asymmetric shorting pins. From the observations above, the impact of one kind of heterogeneity on scanning range varies across different types of elements. This indicates that applying identical heterogeneity to all the elements may not the optimal approach.

Therefore, to further improve the scanning range, the elements in *Array* 3 are modified individually, resulting in *Array* 4. As shown in Fig. 11(a), the *Array* 4 has larger patches loaded on the top of *Type I* elements' asymmetric shorting pins and enhanced back-cavity depth of *Type II* elements. Fig. 11(b) compares the scanning range among *Array*1, 3and *Array* 4, revealing that the *Array* 4 exhibits a further increase in 3-dB gain fluctuation scanning range by 2-10° than the *Array* 3, notably achieving a ±74° scanning range at 29.5 GHz. Thus, applying the SES method and integrating diverse structural modifications across elements within the array can significantly enhance scanning range.

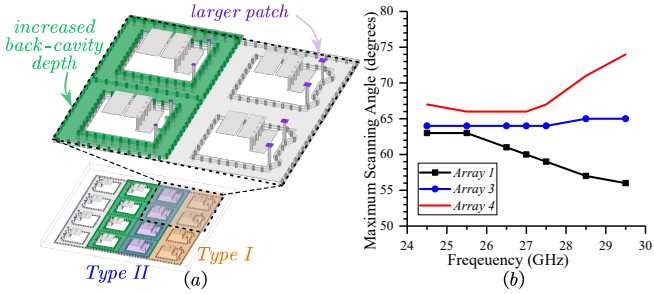

Fig. 11. (a) Configuration of the *Array* 4, (b) 3-dB gain fluctuation scanning range for *Array* 4.

### D. Design Procedures of Heterogeneous Phased Array Using the SES Method

To facilitate understanding of the SES method in heterogeneous phased array design, the procedure is outlined as follows.

1. Design a phased array with traditional element.

2. Create various heterogeneities and apply them to the traditional elements one by one. This generates several heterogeneous arrays, each corresponding to a specific type of heterogeneity.

3. Divide the elements in the array based on their position. For each array, perform the following steps (from a) to c)) to evaluate the effect of a specific heterogeneity on scanning range:

a) Export the AEP for each element, and calculating the S-AEP for each group of element using (9).

b) Extract the 3dB-beamwidth of S-AEP for each group of element obtained in a).

c) Calculate the 3dB-beamwidth difference $\Delta_i$ for each group of element using (10).

4. After the effect of each kind of heterogeneity is known, the rest of procedure is to determine the type of heterogeneity for each group of elements.

As an example of above procedure, as shown in Fig. 12, in this work, a traditional array (*Array* 1) is initially designed. Then, heterogeneities such as the ladder-type metasurface and increased back-cavity depth are applied to *Array*1, resulting in *Array* 2. Subsequently, asymmetrical shorting pins are introduced, producing *Array* 3. Then the array elements are grouped by two types, i.e. Type I and Type II. Then the effect of each heterogeneity on scanning range for these types of elements are investigated. Based on the findings, further modification is applied to *Array* 3, separately for each type of element, obtaining *Array* 4. For instance, it is observed that increasing the back-cavity depth consistently improved the scanning range of Type I elements across the frequency band. Therefore, this modification is implemented in *Array* 4. Similarly, as the asymmetrical shorting pins loading on the Type II elements significantly enhance the scanning range, a larger patch is added to the pins of Type II elements in *Array* 4 to maximize this effect.

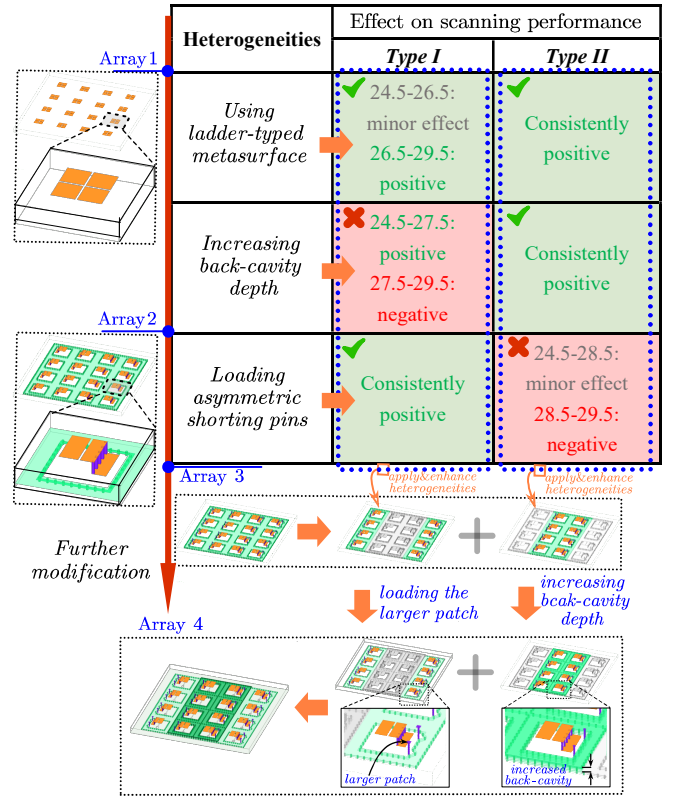

Fig. 12. Design procedure of heterogeneous phased array with the SES method in this work.

However, the active reflection coefficient of the *Array* 4 is not ideal, as shown in Fig. 13. The active reflection coefficient reaches up to -4dB for $\Delta\varphi$=25° and 165°. The next section will address this issue, aiming to reduce the active reflection coefficient.

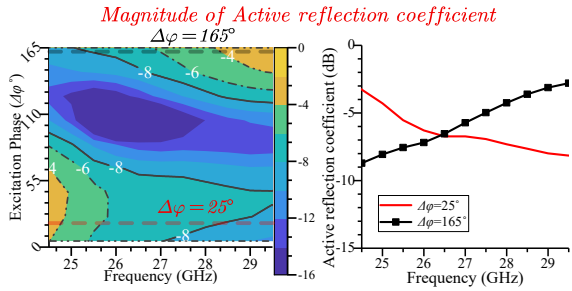

Fig. 13. Magnitude of active reflection coefficient of Port $P_I$ of *Array* 4.

## III. The Active Reflection Self-cancellation Method And Its Application In Reflection Coefficient Reduction For Phased Arrays

As we know, the active reflection coefficient of a phased array often deteriorates when scanning to large angles. To solve this problem, one novel active reflection self-cancellation (ARC) method is proposed. Accordingly, the active reflection coefficient is reduced to below -7.5 dB within the scanning range.

### A. Decoupling Structure Loading for Active Reflection Coefficient Reduction

The active reflection coefficient can be calculated using equation (11). A traditional approach to reduce the active reflection coefficient is by loading a decoupling structure to reduce the coupling coefficient between elements. As shown in Fig. 14, the decoupling structures proposed in [13] are employed to *Array* 4, resulting in the modified *Array* 5.

$$Active S_k = \frac{\sum_n \dot{A}_n S_{kn}}{\dot{A}_k} = \underbrace{S_{kk}}_{\substack{\text{Reflection} \\ \text{component}}} + \underbrace{\sum_{n \neq k} \frac{\dot{A}_n}{\dot{A}_k} S_{kn}}_{\substack{\text{Transmission} \\ \text{component}}} \cdots\cdots (11)$$

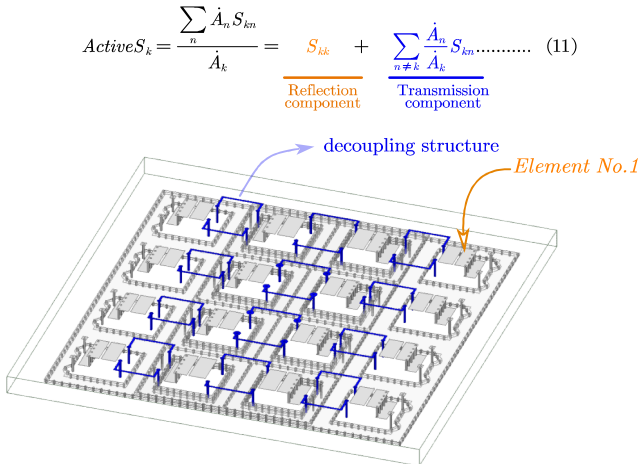

Fig. 14. Configuration of *Array 5* with the decoupling structures.

Fig. 15 shows the S parameters of *Array* 5. As we can see, the isolation between elements in E-plane is improved to over 15 dB, while the reflection coefficients of individual elements are remained below -8.5dB. The active reflection coefficient of *Array* 5 is shown in Fig. 16, performance improvement is observed when $\Delta\varphi = 25°$. However, for $\Delta\varphi = 165°$, the active reflection coefficient remains poor. Noting that, although the decoupling structure enhances the isolation between elements, it affects the phase of transmission coefficient and increases the reflection coefficient. When $\Delta\varphi = 165°$, the transmission and reflection component are nearly in phase. Therefore, loading the decoupling structure can hardly reduce the active reflection coefficient for large-angle phase excitations.

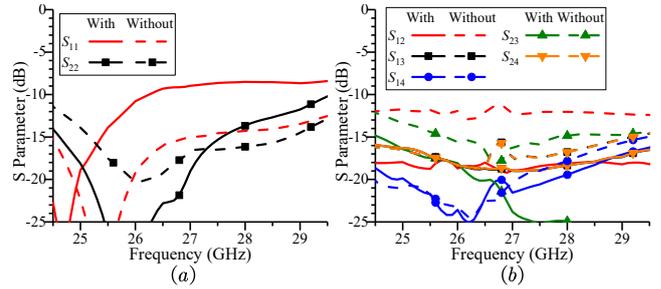

Fig. 15. (a) Reflection and (b) Transmission coefficient before and after loading decoupling structures.

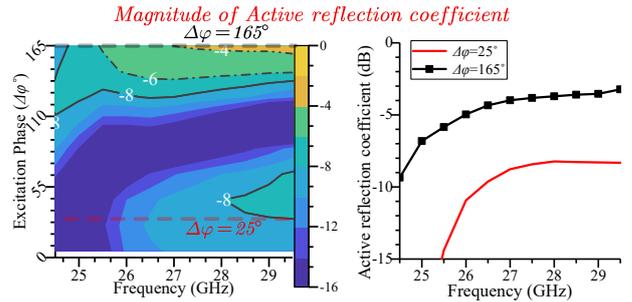

Fig. 16. Magnitude of active reflection coefficient of Port $P_I$ for *Array 5*.

### B. The ARC Method for Active Reflection Coefficient Reduction

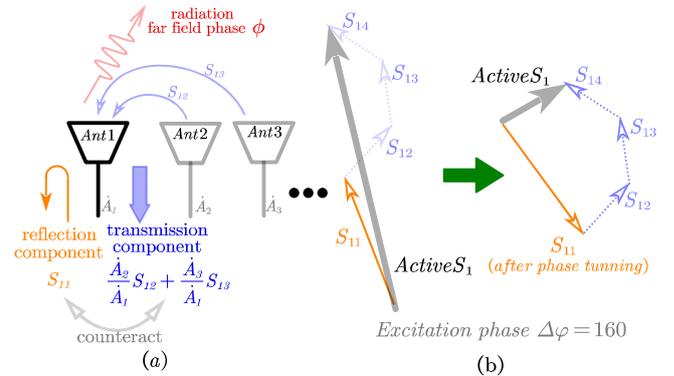

Fig. 17. Sketch for (a) Transmission and reflection components of active reflection coefficient, (b) Vector composition of components in ARC method.

To reduce the active reflection coefficient of a phased array during large-angle scanning, the ARC method is proposed. The principle of this method is sketched in Fig. 17(a), the active reflection coefficient in (11) can be decomposed into the reflection component of element and the transmission component from the other elements. Therefore, as shown in Fig. 17(b), the active reflection coefficient can be reduced by adjusting the phase of the reflection component to make it cancel out with the transmission component.

The ARC method requires flexible tuning of the element reflection coefficient phase without significantly altering its transmission phase to free space. To this end, two particular approaches are presented. As the phase of the reflection coefficient at a specific frequency $\omega_0$ corresponds to the angle of the vector on the Smith chart, as shown in Fig. 18, which points from the center to the position on the impedance loci. Therefore, adjusting the reflection coefficient phase essentially involves rotating the impedance loci relative to the center on the smith chart.

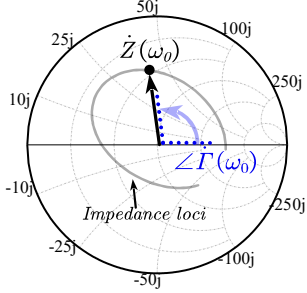

Fig. 18. Relationship between impedance loci $\dot{Z}(\omega)$ and phase of reflection coefficient $\angle \dot{\Gamma}(\omega)$.

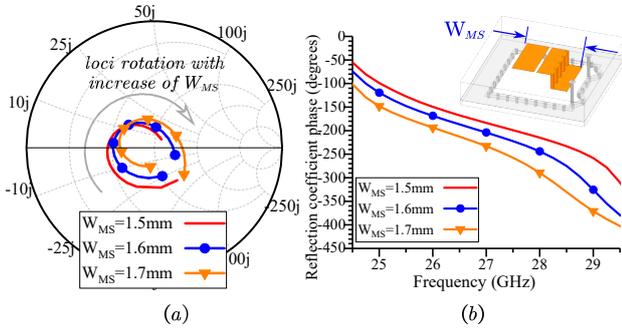

Fig. 19. (a) Reflection coefficient phase and (b) Impedance loci for element with different width $W_{MS}$.

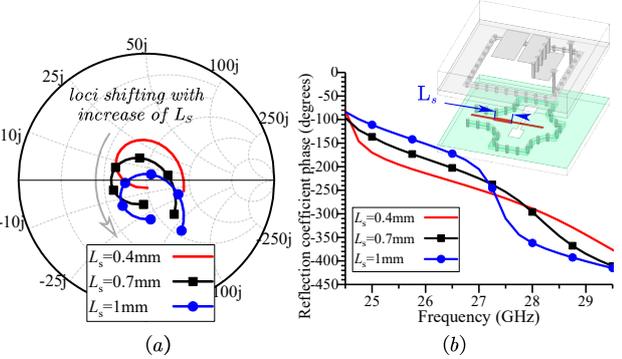

Fig. 20. (a) Reflection coefficient phase and (b) Impedance loci for element with different stepped-impedance line length $L_S$.

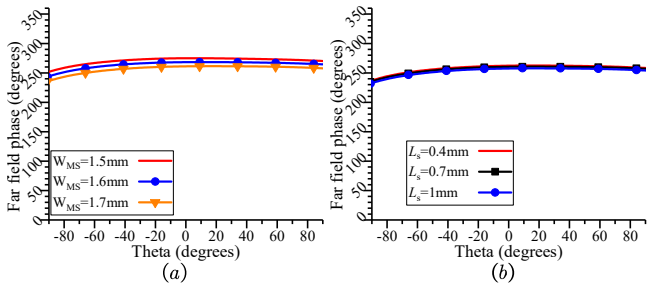

Fig. 21. Far field phase for elements with different (a) Width $W_{MS}$ and (b) Stepped-impedance line length $L_S$.

On the one hand, this can be achieved by shifting the resonant frequency, which is implemented by adjusting the length of the metasurface. As shown in Fig. 19(a), changing the metasurface length alters the element resonances, which in turn rotates the impedance loci on the Smith chart. This rotation results in either an increase or decrease in the reflection coefficient phase across the whole frequency band. As shown in Fig. 19(b), when the metasurface length $W_{MS}$ increases from 13.5 mm to 15.5 mm, the reflection coefficient phase decreases by approximately 50°. On the other hand, the reflection coefficient phase can be adjusted by shifting the impedance loci. This can be achieved by introducing stepped impedance lines. As shown in Fig. 20(a), increasing the length of the stepped impedance line $L_S$ shifts the impedance loci to the capacitive region. Consequently, as illustrated in Fig. 20(b), the reflection coefficient phase increases by 63° from 24.5 GHz to 27.36 GHz, while it decreases by 71° from 27.5 GHz to 30 GHz. Additionally, note that the far field phase should remain constantly to guarantee an unchanged radiation pattern of the array. Fortunately, the far field phase deviates slightly during the reflection coefficient phase tuning with these two methods. As shown in Fig. 21, the far field variation is less than ±7° for resonance tuning and ±2.5° for the impedance loci position tuning. This indicates that altering the reflection coefficient phase affects little on beam direction.

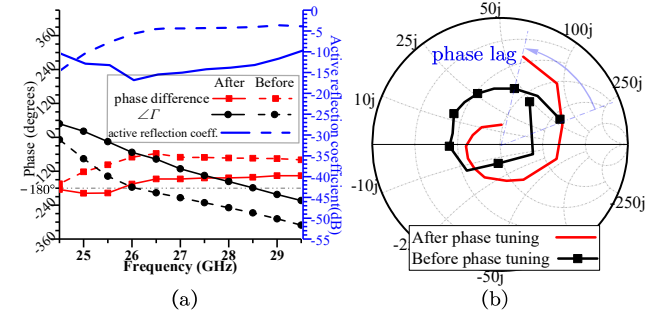

Fig. 22. Performance of the *Element No.1* before and after phase tuning: (a) Phase difference between reflection and transmission component, reflection coefficient phase (left) and active reflection coefficient (right), (b) Impedance loci.

The ARC method is then employed to *Array* 5 to reduce its active reflection coefficient. For clear demonstration, the active reflection coefficient for the *Element No.1* when $\Delta \varphi = 165°$ is reduced by adjusting the metasurface length. As shown in Fig. 22(a), the active reflection coefficient of *Element No.1* exceeds -4dB initially, with phase difference between the transmission component and $\angle \dot{\Gamma}$ near -74°. Therefore, an increment in $\angle \dot{\Gamma}$ can facilitate self-cancellation of the active reflection coefficient. To this end, the length of *Element No.1* is reduced from 1.9mm to 1.55mm, resulting in an anti-clockwise rotation in its impedance loci and thus a phase decrement for around 70°. As a result, the phase difference between the transmission component and $\angle \dot{\Gamma}$ is decreased to around -140°, and the active reflection coefficient is reduced to below -9.7dB.

Similar procedure is also applied to the other elements in *Array 5* to adjust their reflection coefficient phase. Consequently, the active reflection coefficient seen behind each element is reduced. Note that a reduction in the element active reflection coefficient also result in a corresponding reduction in the active reflection coefficient observed behind the power divider. Therefore, the active reflection coefficient for the power divider port $P_1$-$P_4$ within the scanning range, which in accord with excitation phase $|\Delta\varphi| \lesssim 165°$, is finally reduced to below -7.5 dB, as shown in Fig. 23. Therefore, the proposed ARC method effectively reduces the active reflection coefficient when the phased array scans to large angles.

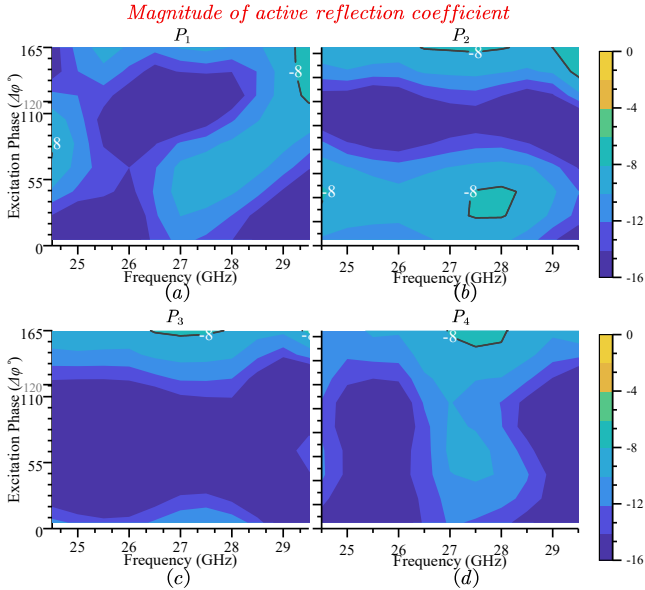

Fig. 23. Magnitude contour plot of active reflection coefficient of Array5 after phase tuning for (a) $P_1$, (b) $P_2$, (c) $P_3$, (d) $P_4$.

## IV. RESULTS AND DISCUSSION

### A. Configuration of the Proposed Phased Array

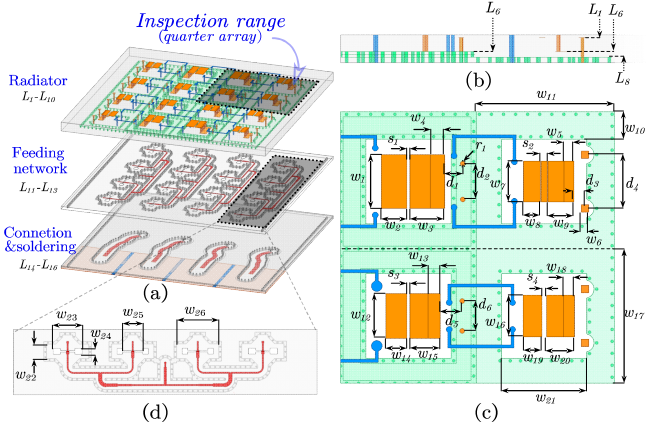

Fig. 24. Configuration of prototype for manufacture: (a) 3D view, (b) Side view of inspection range, (c) Top view of inspection range, (d) Configuration of the feeding structure.

For verification of above design strategies, *Array 5* is fabricated and measured. The detailed configuration is illustrated in Fig. 24. The array is manufactured with LTCC process, where $L_1$-$L_{10}$ are the radiator layers, $L_{11}$-$L_{14}$ are the feeding network layers, $L_{13}$-$L_{16}$ are layers for soldering with connectors. Note that only dimensions for quarter of array (the inspection range) are provided due to symmetry. The key dimensions are listed in the Table. I.

TABLE I
DIMENSIONS OF THE PROPOSED ARRAY. (UNITS: MM)

| $w_1$ | $w_2$ | $w_3$ | $w_4$ | $w_5$ | $w_6$ | $w_7$ | $w_8$ | $w_9$ | $w_{10}$ |
|---|---|---|---|---|---|---|---|---|---|
| 1.8 | 0.85 | 1.15 | 0.48 | 0.33 | 0.25 | 1.37 | 0.56 | 0.86 | 0.93 |
| $w_{11}$ | $w_{12}$ | $w_{13}$ | $w_{14}$ | $w_{15}$ | $w_{16}$ | $w_{17}$ | $w_{18}$ | $w_{19}$ | $w_{20}$ |
| 4.5 | 1.5 | 0.4 | 0.7 | 1 | 1.4 | 4.5 | 0.36 | 0.63 | 0.93 |
| $w_{21}$ | $w_{22}$ | $w_{23}$ | $w_{24}$ | $w_{25}$ | $w_{26}$ | $d_1$ | $d_2$ | $d_3$ | $d_4$ |
| 2.83 | 1.02 | 2 | 0.4 | 1 | 2.92 | 0.6 | 1.2 | 0.37 | 1.8 |
| $d_5$ | $d_6$ | $r_1$ | $s_1$ | $s_2$ | $s_3$ | $s_4$ | $L_n$ | | |
| 0.73 | 1 | 0.08 | 0.1 | 0.25 | 0.1 | 0.15 | 0.094×n | | |

### B. Measured Results

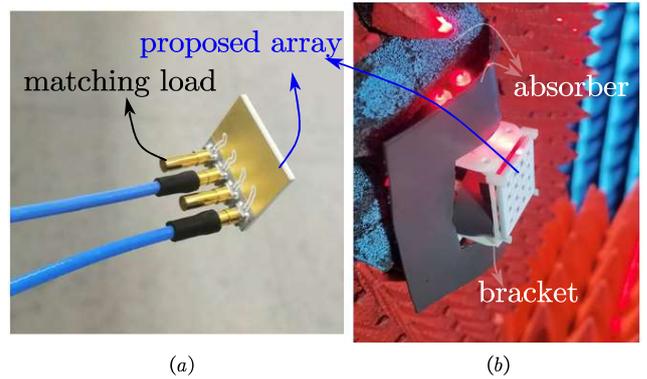

Fig. 25. Measurement configuration for: (a) S parameter and (b) Radiation performance.

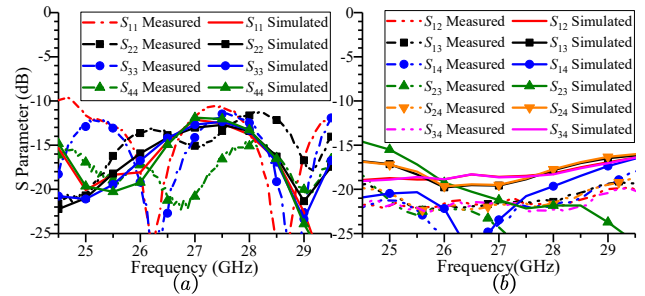

Fig. 26. Measured S parameters: (a) Reflection coefficient, (b) Transmission coefficient.

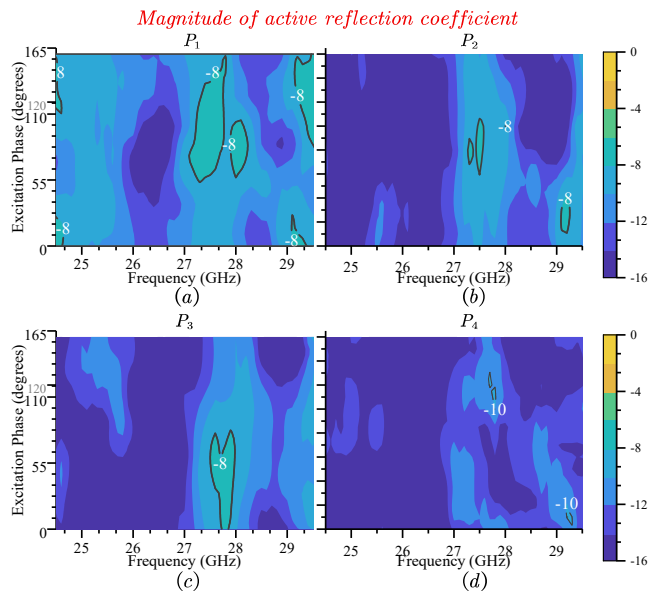

TABLE II
COMPARISON AMONG SEVERAL PHASED ARRAYS

| Works | Operating band (GHz) | Maximum scanning range# | Active reflection coefficient (dB) | Array scale | Peak gain (dBi) | Method to achieve wide-angle scanning | Method to reduce active reflection coefficient |
|---|---|---|---|---|---|---|---|
| [12]* | 24.5-29.5 | ±62° | ≤-2.8 | 1×4 | 14.7 | pseudo-curved surface conformal | None |
| [10]* | 24.5-29.5 | ±72° | ≤-3.2 | 4×4 | 11.8 | pseudo-curved surface conformal | None |
| [4] | 2.55-3.12 | ±70° | ≤-5 | 1×8 | 13.7 | wide-beamwidth element | parasitic structure loading |
| [2] | 4.4-5.05 | ±65° | ≤-7.5 | 1×8 | 14.35 | wide- beamwidth element | nodal lines loading |
| **This work** | **24.5-29.5** | **±74°** | **≤-7.5** | **4×4** | **14.4** | **SES method** | **ASC method** |

*The active reflection coefficient is obtained from the simulated result of the model provided by the article.
# The 3-dB gain fluctuation scanning range is compared.

Fig. 27. Magnitude contour plot of active coefficient calculated from the measured S parameters: (a) $P_1$, (b) $P_2$, (c) $P_3$, (d) $P_4$.

The S parameters of array were measured with ROHDE&SCHWARZ ZVA40 vector analyzer. As shown in Fig. 25(a), all the other ports are terminated with matching load except for those under measured. As can be observed, the measured isolations are consistent with the simulated one. However, some difference exists between the measured and simulated reflection coefficients, which is due to soldering and connector transitions, nevertheless, all of the reflection coefficients are below than -10dB. Then the active reflection coefficient is calculated based on the measured S parameters. Consequently, as shown in Fig. 27, good performance lower than -7.5 dB is observed, despite some difference is observed between the calculated results and the simulated result in Fig. 23, which can stem from magnitude errors in reflection coefficient and phase differences introduced by multiple plugging and unplugging of connectors.

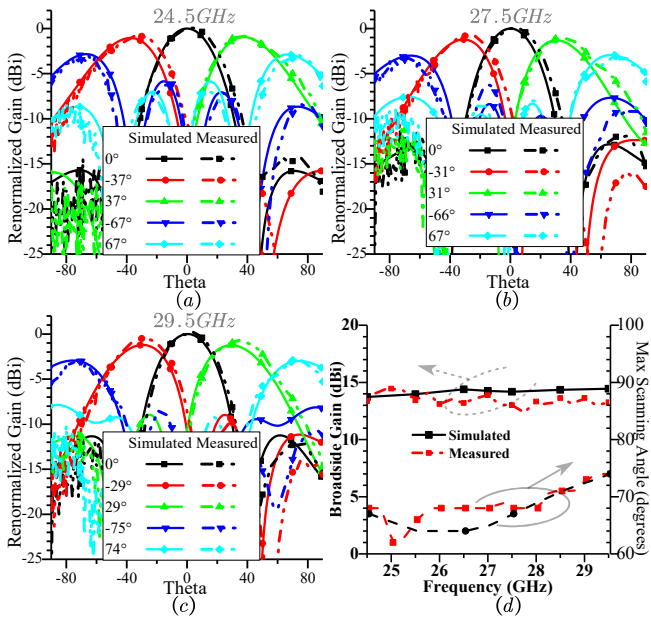

Fig. 28. Measured radiation performance and scanning range: normalized radiation patterns for different scanning angles at (a) 24.5GHz, (b) 27.5GHz, (c) 29.5GHz. (d) Broadside gain when not scanning (left) and 3-dB gain fluctuation scanning range within band (right).

The active element patterns (AEP) of each port of the array was measured in a compact antenna test range chamber, as shown in Fig. 25(b). Note that the scanning beam is calculated based on the measured AEPs. Consequently, the normalized scanning patterns at 24.5, 27.5, and 29.5 GHz are shown in Fig. 28(a-c), good agreement is observed. Fig. 28(d) illustrates the broadside gain when the array is not scanning, along with the 3-dB gain fluctuation scanning angle within the band. The measured and simulated results are consistent.

### C. Discussions

For further demonstration, a comparison is conducted among several phased arrays that of state-of-the-art [2], [4], [10], [12], as shown in Table II. The array in [12] and [10] employs pseudo-conformal technology to enhance scanning range. However, the design method for heterogeneous elements merely simulates the tilted beams of elements in a conformal surface array, this qualitative design strategy limits the scanning range of the array. Additionally, the poor active reflection coefficient at large scan angles in these works also reduce the realized gain when scanning to large angle, thus hinges the scanning range. The arrays in [4] and [2] achieved wide-angle scanning with wide-beam elements. Meanwhile, the active reflection coefficient was reduced by loading parasitic structures and employing a neutral line decoupling structure in these two arrays. However, the active reflection coefficient in [4] is still not ideal, which is -5dB, and the scanning range in [2] is limited by the wide-beam element's beamwidth. In contrast, the array proposed in this work features wide scanning range up ±74° by applying the SES method. Additionally, thanks to the ARC method, the active reflection coefficient is reduced to below -7.5dB. Therefore, the proposed array features a wide scanning range and good active reflection coefficient, indicating one feasible solution for millimeter-wave wide-angle scanning phased arrays.

### V. CONCLUSIONS

In this work, a novel SES method is introduced to improve the scanning range of a heterogeneous array. The quantitative analysis of how different types of heterogeneity contribute to scanning range is described. Additionally, an ARC method is proposed to reduce reflection coefficient and realized gain drop in case of wide-angle scanning. By combining these two methods, a phased array with customized heterogeneity for each element is developed, achieving a wide scanning range of up to ±74° and a reflection coefficient lower than -7.5 dB.